\documentclass[prl,superscriptaddress,floatfix,letterpaper, twocolumn]{revtex4}

\usepackage[utf8]{inputenc}

\usepackage{natbib}
\usepackage{amsmath}
\usepackage{amssymb}

\usepackage{graphicx}

\usepackage{verbatim}

\newcommand{\ket}[1]{| #1 \rangle}

\usepackage[urlcolor=blue, hyperindex, colorlinks, bookmarks=true]{hyperref}

\begin{document}

\title{Coherent Control of a Superconducting Qubit with Dynamically Tunable Qubit-cavity Coupling}

\author{A.J. Hoffman}
\author{S.J. Srinivasan}
\affiliation{Department of Electrical Engineering, Princeton University, Princeton, NJ 08544, USA}
\author{J.M. Gambetta}
\affiliation{IBM T.J. Watson Research Center, Yorktown Heights, NY 10598, USA}
\author{A.A. Houck}
\affiliation{Department of Electrical Engineering, Princeton University, Princeton, NJ 08544, USA}

\date{August 12, 2011}

\ifpdf
\DeclareGraphicsExtensions{.pdf, .jpg, .tif}
\else
\DeclareGraphicsExtensions{.eps, .jpg}
\fi

\begin{abstract}
We demonstrate coherent control and measurement of a superconducting qubit coupled to a superconducting coplanar waveguide resonator with a dynamically tunable qubit-cavity coupling strength.  Rabi oscillations are measured for several coupling strengths showing that the qubit transition can be turned off by a factor of more than 1500.  We show how the qubit can still be accessed in the off state via fast flux pulses.  We perform pulse delay measurements with synchronized fast flux pulses on the device and observe $T_1$ and $T_2$ times of 1.6 and 1.9 $\mu$s, respectively.  This work demonstrates how this qubit can be incorporated into quantum computing architectures.
\end{abstract}


\maketitle

Circuit quantum electrodynamics (cQED) is a promising architecture for a quantum computer based on superconducting qubits~\cite{Blais2004,Wallraff2004,Girvin2009}.  This architecture, in which qubits are strongly coupled to a transmission line resonator, provides multiplexed control and readout, as well as photon-mediated qubit-qubit interactions needed for multiqubit gates~\cite{Majer2007,Sillanpaa2007}.  Typically, both single and multiqubit gates are turned off by detuning the qubits from microwave drives and from other qubit resonances~\cite{Majer2007,Sillanpaa2007,Lucero2008,DiCarlo2009, Yamamoto2010}.  However, with the strong interactions needed for fast gates, the detuning needed for low error rates can be quite large, resulting in problems of spectral crowding and errors from residual coupling. Several approaches have been taken to tunable coupling, both for direct coupling between qubits and in a cQED architecture~\cite{Allman2009,Bialczak2011,Filipp,Galiautdinov,Harris2007,Hime2006,Liu2006,Niskanen2007,Ploeg2007}.  Here, we demonstrate coherent operation of a tunable coupling qubit (TCQ), in which both the qubit-cavity coupling strength and frequency of the qubit can be tuned independently, thus allowing unwanted couplings to be turned off more effectively.

In previous work, we introduced the TCQ and showed how it interacted resonantly with a superconducting cavity~\cite{Srinivasan2011}.  This qubit consists of two transmons directly coupled by a large capacitor which determines the characteristic interaction energy, $\hbar J$~\cite{Srinivasan2011, Gambetta2011}.  By tuning the two lowest transmon-like energy levels into resonance, the lowest excitation of the collective system becomes the antisymmetric superposition of the single transmon excitations, which has no net coupling to the cavity.  By independently tuning the two transmon levels, both the frequency and coupling of the collective device can be independently tuned~\cite{Gambetta2011}.  The coupling strength can be changed adiabatically, compared with $J$, from zero coupling to the strong coupling regime by tuning these levels into and out of resonance with one another.  An energy level diagram of the \emph{hybridized states}, without any shifts due to the cavity, is shown in Fig. 1.  In this work, we use the states $\ket{00}$ and $\ket{\widetilde{10}}$ with a transition energy of $\hbar \omega_{10-00}$ as the logical states of our qubit.  Both this transition and the $\ket{\widetilde{20}}-\ket{10}$ transition with energy $\hbar \omega_{20-10}$ have low coupling strengths when the bare, transmon-like energy levels are in resonance, $g_{10-00}$ and $g_{20-10}$, respectively.  Conversely, the $\ket{\widetilde{01}}-\ket{00}$ and $\ket{\widetilde{11}}-\ket{\widetilde{10}}$ transitions, with transition energies, $\hbar \omega_{01-00}$ and $\hbar \omega_{11-10}$, have high coupling strengths.

In this paper, we operate the qubit in the dispersive regime and demonstrate coherent control while changing the coupling strength on time scales suitable for single qubit gate operations.  In the space spanned by two flux control lines, we identify a contour of constant dressed qubit frequency.  Moving along this contour, we demonstrate that the qubit-cavity interaction can be turned off by more than a factor of 1500, and that the qubit transition cannot be driven when this coupling is off.  Moreover, we can dynamically turn the interaction back on and control the qubit using synchronized fast flux pulses and rf control pulses.  This work lays the foundation for the practical use of this device in quantum systems.

The superconducting charge qubit used in these experiments is nearly identical to the qubit reported in Ref.~\cite{Srinivasan2011}.  The device is fabricated on a sapphire substrate using electron beam lithography and a double-angle Al evporation.   The geometry consists of two islands each connected to a third, common island via a pair of split Josephson junctions in a SQUID loop.  The TCQ is fabricated in a notch between the center pin and ground plane of a superconducting Nb coplanar waveguide resonator with a bare resonance frequency of 9.54 GHz and a $Q$ of 470.  The sample box is encased in Ecosorb CR-124 and then cooled to 20 mK in a dilution refrigerator~\cite{Corcoles}.  Two fast flux bias lines connected to separate low-noise current sources control the flux through the two SQUID loops and are used to change the effective energy of the split junction pairs.  Applying the appropriate flux through the SQUID loops allows independent control of the coupling strength $g_{10-00}$ and qubit frequency $\omega_{10-00}$.

In this work, we are mainly concerned with changing only the coupling strength of the qubit to the cavity while keeping the qubit frequency fixed.  Since the flux controls allow for a wide range of coupling strengths and dressed qubit transition frequencies, it is necessary to find the control subspace that corresponds to constant dressed qubit frequency.  This subspace accounts for any dispersive shifts due to changes in qubit-cavity coupling.  To accomplish this, we use standard dispersive readout techniques of cQED:  monitoring the amplitude and phase of cavity transmission while applying a second spectroscopy tone.  Here, though, we keep the spectroscopy tone at a constant frequency of $7.500\,\mathrm{GHz}$ while sweeping the two control fluxes.  When the dressed qubit frequency, which is a function of the two control fluxes, is resonant with the $7.500\,\mathrm{GHz}$ spectroscopy tone, a change in the cavity transmission is measured~\cite{Wallraff2004}.  Over a wide range of control voltages, it is then possible to extract a contour that corresponds to where the dressed qubit frequency is $7.500\,\mathrm{GHz}$; such a contour is shown in Fig. \ref{figure2}a.

\begin{figure}
	\includegraphics[width=0.47\textwidth,clip]{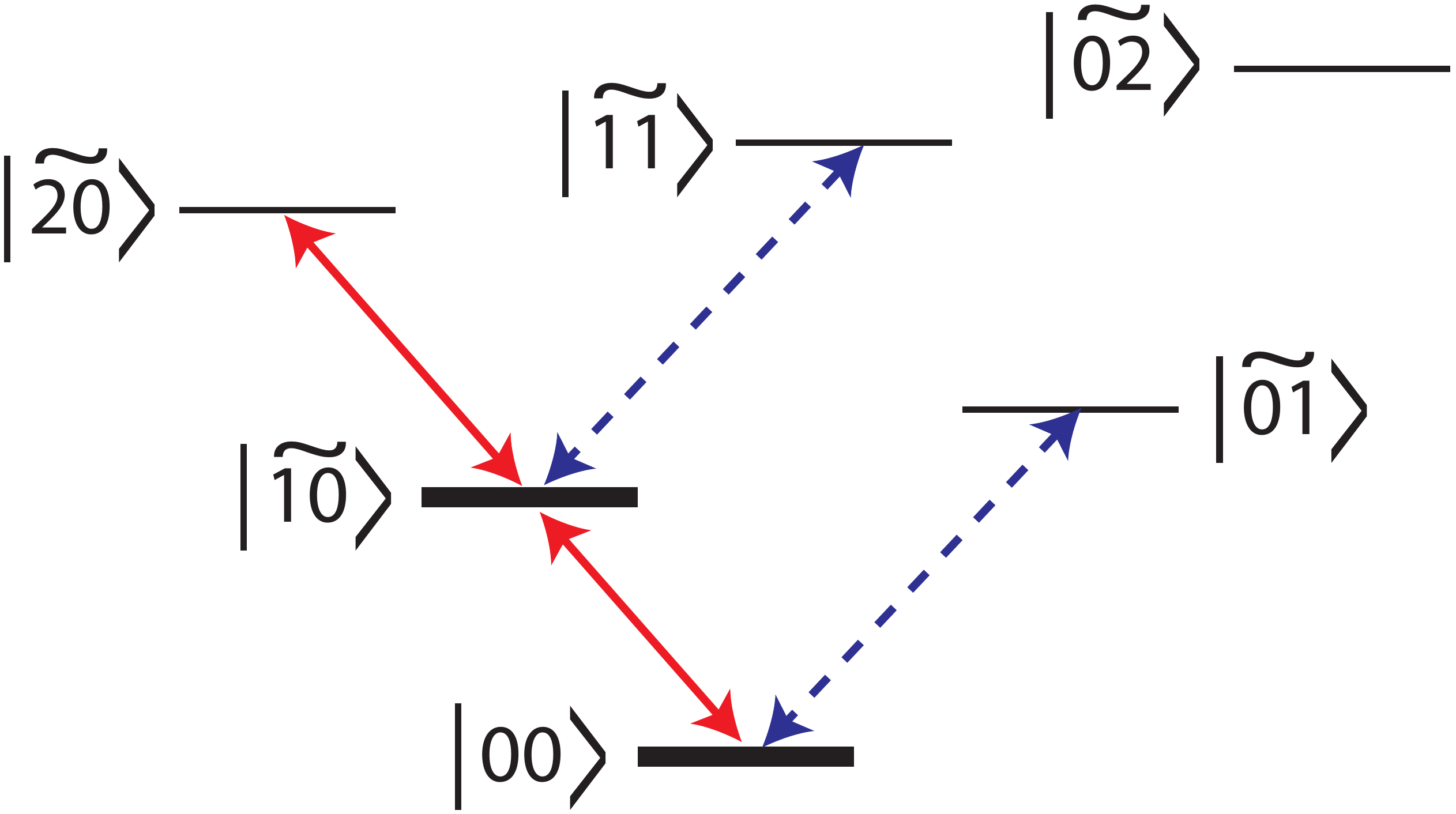}
	\caption{\label{figure1}  Energy level diagram for the TCQ showing the hybridized energy levels.  The transitions that have a high probability of occurring are indicated by arrows. Considering that the system is primarily in the $\ket{00}$ or $\ket{\widetilde{10}}$ states, single-photon transitions leading out of these two states have the maximum transition probabilities and they are indicated by arrows. The red, solid arrows indicate transitions with low coupling strengths and the blue, dashed arrows indicate transitions with high coupling strengths.  The levels shown here are for the bare energy levels of the device; there are no effects of coupling to a cavity. In this work, the $\ket{00}$ and $\ket{\widetilde{10}}$ states are used as the logical states of the qubit.}
\end{figure}

Along this contour of constant qubit frequency, the qubit-cavity coupling strength, $g_{10-00}$, changes due to the quantum interference of the two transmon-like halves of the TCQ.  In Fig. \ref{figure2}b, we measure the frequency response of the qubit while moving along the parameterized contour and can clearly see that the dressed qubit frequency remains $7.500\,\mathrm{GHz}$.  Moreover, in this constant power measurement, the amplitude of the response is related to the coupling strength between the qubit and the superconducting resonator.  When the coupling is small, little response is seen because the qubit cannot be driven.  The disappearance of a signal corresponds to the situation where the qubit-cavity coupling is tuned through zero.

\begin{figure}
	\includegraphics[width=0.47\textwidth,clip]{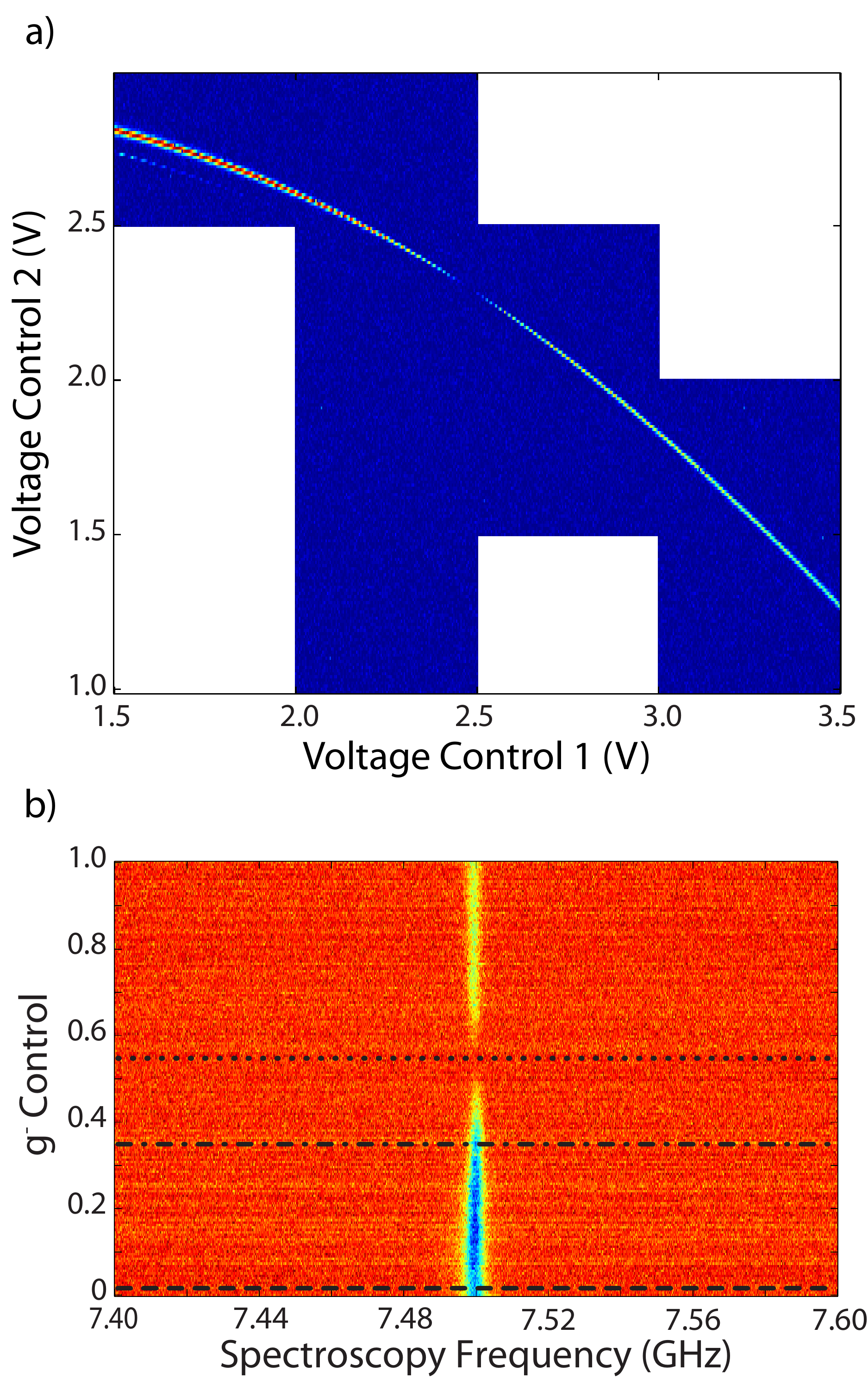}
	\caption{\label{figure2} (a) Observed cavity transmission versus the two control voltages with a fixed spectroscopy tone at $7.5 \mathrm{GHz}$.  Both the dressed qubit frequency and coupling strength are functions of the control voltages.  The contour shows where the qubit is resonant with the $7.5 \mathrm{GHz}$ tone and is therefore driven between the ground and excited states.  (b) Measured dressed frequency response of the qubit while moving along the $7.5\,\mathrm{GHz}$ contour in Fig. 1.  The dressed qubit frequency remains constant at $7.5\,\mathrm{GHz}$.  The amplitude of the response is related to the coupling strength between the qubit and the superconducting resonator.  The point where the signal disappears corresponds to coupling strengths where the qubit cannot be driven by the spectroscopy tone.  The dotted and dashed lines indicate $g_{10-00}$ control values where the measurements were performed for Fig. 3.}
\end{figure}

\begin{figure}
	\includegraphics[width=0.47\textwidth,clip]{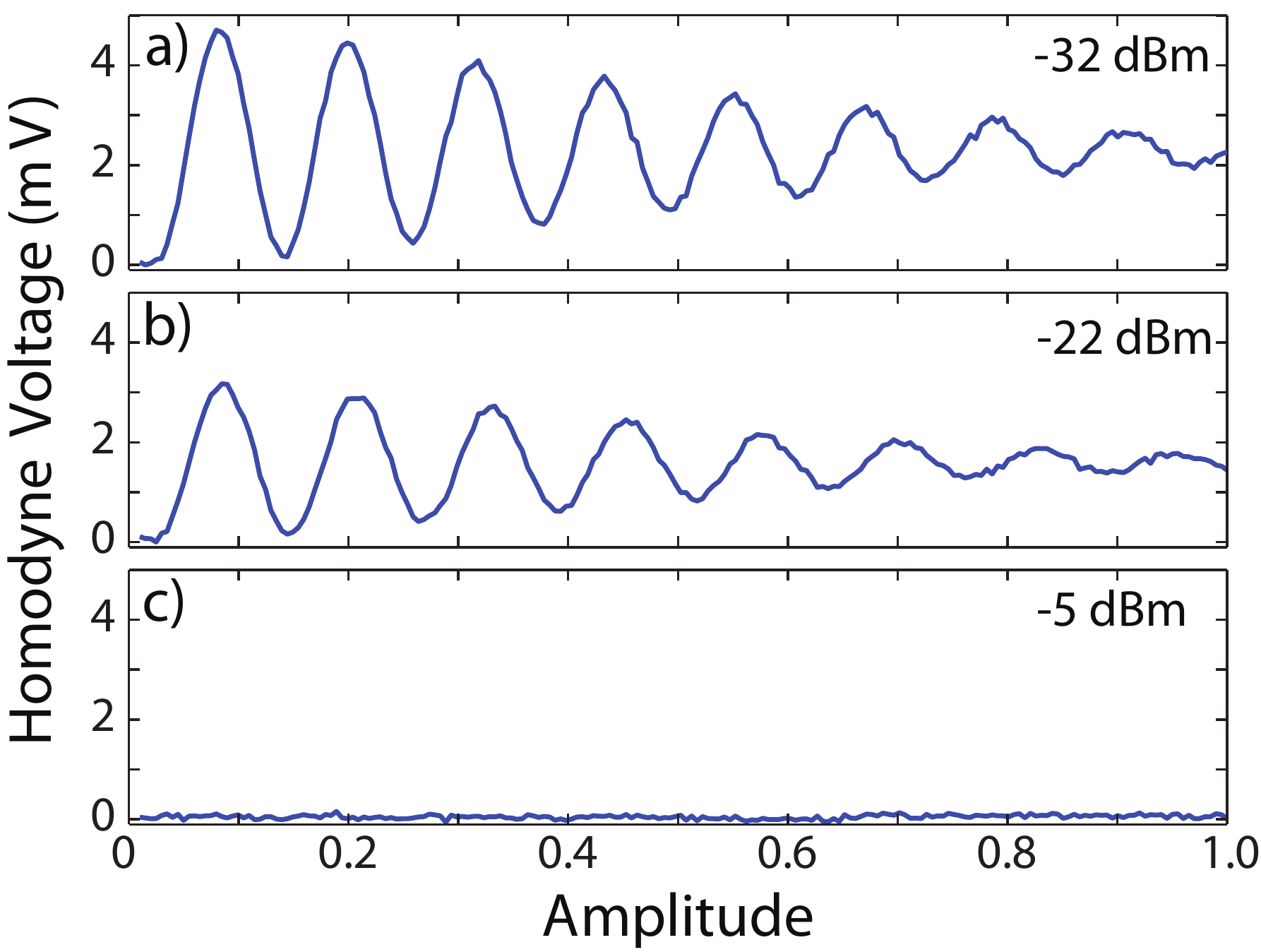}
	\caption{\label{figure3} Rabi oscillations for three different qubit-cavity coupling strengths and a fixed dressed qubit frequency of $7.5\,\mathrm{GHz}$.  Panels (a), (b), and (c) correspond to the dashed, dot-dash, and dotted lines in Fig. 2, respectively.  In (a), a spectroscopy power of -32 dBm is used.  To keep the number of oscillations approximately the same for the lower qubit-cavity coupling strength in (b), the spectroscopy power is increased to -22 dBm.  In panel (c), 27 dBm more power than that in (a) is applied and no oscillations are observed.  Given the measurement noise, we put a bound of $1/10$ of a Rabi oscillation.}
\end{figure}

  Time domain measurements provide a more quantitative assessment of any residual coupling at the $g_{10-00}=0$ point.  The rate of Rabi driving is proportional to the coupling strength $g_{10-00}$ and the applied drive amplitude as per the equation $\Omega_{Rabi} = g_{10-00} \sqrt{n}$, where $n$ is the number of drive photons~\cite{Blais2004}.  In Fig \ref{figure3}, we demonstrate Rabi oscillations at three different points on the constant frequency contour; these three points are marked on Fig. \ref{figure2}b.  Figure \ref{figure3}a, b show Rabi oscillations at high and medium coupling respectively.  In the two panels, the oscillation rate is kept nearly constant by increasing the applied rf spectroscopy power by 10 dB to compensate for the reduction in qubit-cavity coupling.  Fig. \ref{figure3}c shows the measurement at the $g_{10-00}=0$ point, with 27 dB more rf power than at the high coupling point.  No excitation is visible.  Given the measurement noise, we should easily be able to detect a tenth of a Rabi oscillation; that we see no excitation puts a lower bound on the change in the Rabi rate of a factor of 80.  Together with the much higher excitation power, we estimate that the coupling is at least 1500 times smaller at the $g_{10-00}=0$ point compared with the high coupling point.  If several qubits were in a single cavity, this tuning provides protection against single qubit gate errors in one qubit while a second qubit is driven.

Statically decoupling qubits from the microwave cavity is of little use if the coupling strength cannot be dynamically increased when Rabi driving is desirable.  To this end, we utilize fast flux pulses, created using two analog channels of a Tektronix 5014 arbitrary waveform generator, to coherently control qubits that have a rest bias at the $g_{10-00}=0$ point.  We apply synchronized 60 ns flux bias pulses to each bias line during which the qubit can be driven.  To ensure that there are no slow transients when the qubit is returned to its off state, we apply an additional flux pulse of the opposite sign, so that the total integrated flux on each line is zero~\cite{Johnson2011}.  These techniques wre used to move the qubit from a rest bias where $g_{10-00} = 0$ to a point of higher coupling for short periods of time in order to excite the qubit.  After the excitation, the qubit is returned to the $g_{10-00}=0$ point and qubit state readout is performed, resulting in the Rabi oscillations shown in Fig. 4a.

\begin{figure}
	\includegraphics[width=0.47\textwidth,clip]{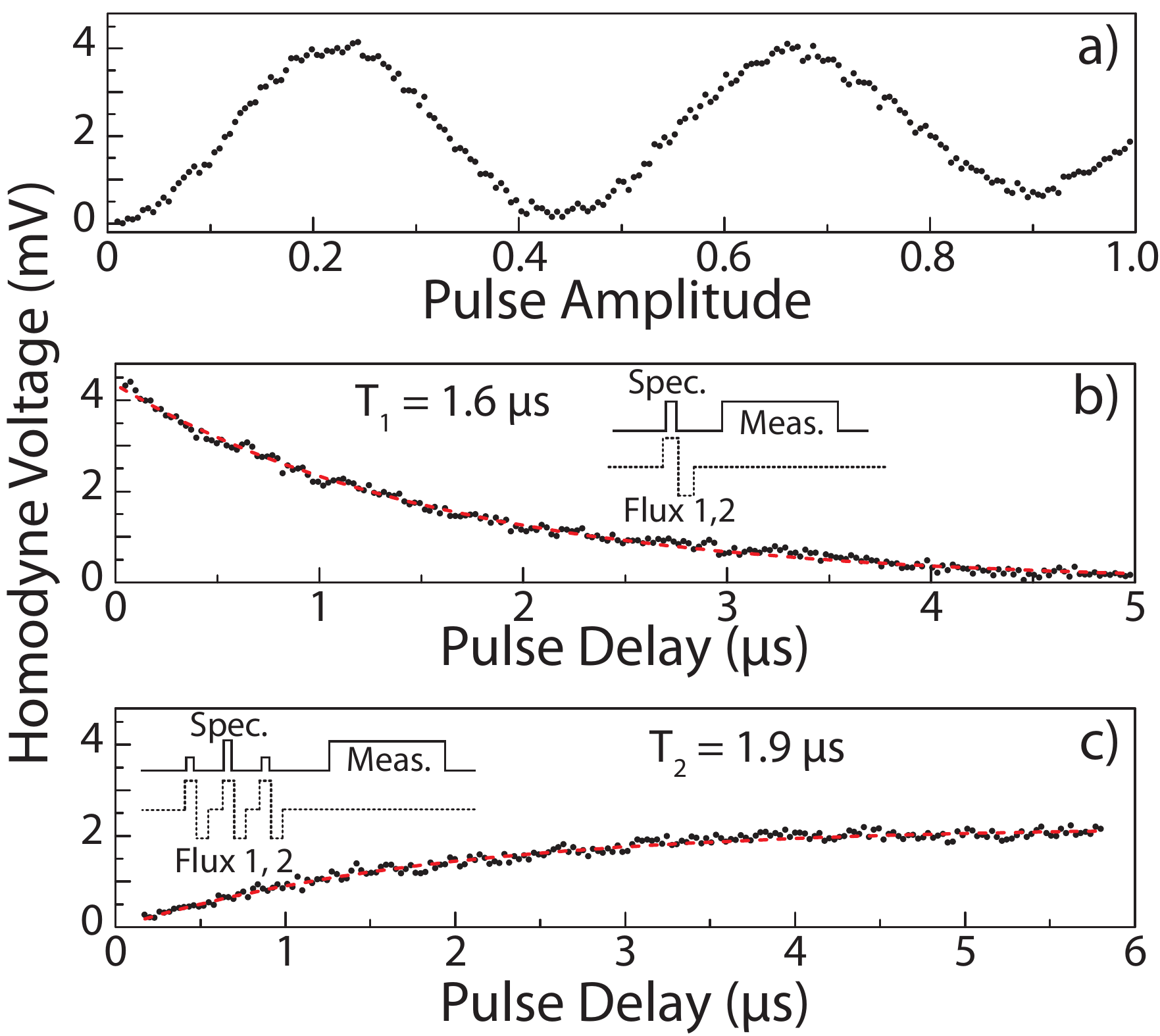}
	\caption{\label{figure4} (a) Observed Rabi oscillations when the qubit starts in the $g_{10-00} = 0$ state and is simulataneously moved to a large $g_{10-00}$ state and driven by a $7.5\,\mathrm{GHz}$ spectroscopy pulse of varying amplitude.  The fast flux pulse is 60 ns in duration and is followed by an identical pulse of the opposite sign so that the total pulse integral is zero; these zero integral pulses help reduce slow transients.  (b) Pulsed measurements showing the probability of the qubit being in the excited state as a function of delay following a $\pi$-pulse.  The qubit starts in the $g_{10-00}=0$ state and is excited with a $\pi$-pulse in the manner described in (a); a pulsing scheme is included as an inset to the figure.  The measured $T_1$ is 1.6 $\mu$s.  (c)  Hahn echo measurements with the qubit starting in the $g_{10-00} = 0$ state.  Each of the pulses in the Hahn sequence is synchronized with a pair of fast flux pulses.  A pulsing scheme is included as an inset to the figure.  The measured $T_2$ time is 1.9 $\mu$s. }
\end{figure}

Using these fast flux bias pulses, we first measure $T_1$ by applying a $\pi$-pulse that is synchronized with the fast flux pulse, and measure the qubit excitation probability after a delay.  We measure $T_2$ using a Hahn echo experiment.  The qubit is returned to the $g_{10-00}$ state after each pulse in the Hahn echo sequence.  The results of these measurements and the pulse schemes are shown in Fig \ref{figure4}b, c.  The measured $T_1$ and $T_2$ times are $1.6$ and $1.9\,\mathrm{\mu s}$, respectively.  The times are only slightly shorter than the $1.9$ and $2.8\,\mathrm{\mu s}$ times recorded at high $g_{10-00}$ without any fast flux pulses.

The dispersive shift, $\chi$, on the cavity’s resonance frequency is dependent on the various energy levels in the system and their corresponding transition dipoles. Ignoring contributions from transitions with low probability, the difference in the cavity transmission frequencies when the qubit is in the ground or excited state, $2\chi$, can be approximated as $2\chi \approx \frac{2g_{10-00}^2}{\Delta_{10-00}} - \frac{g_{20-10}^2}{\Delta_{20-10}} + \frac{g_{01-00}^2}{\Delta_{01 -00}}  - \frac{g_{11-10}^2}{\Delta_{11-10}}$, where $\Delta_{i-j} = \hbar(\omega_{ij} - \omega_{r})$ and $\omega_r$ is the resonator frequency. While the magnitude of both $g_{10-00}$ and $g_{11-10}$ can be extremely small, the other couplings do not vanish at the same point. In fact, $g_{01-00}$ and $g_{11-10}$ are large because when the dipole coupling of the antisymmetric state goes to zero, the coupling of the symmetric state is at its maximum value~\cite{Srinivasan2011}. However, due to the interaction between the two independent transmon levels, $\Delta_{01 -00} \neq \Delta_{11-10}$. These nonvanishing terms are accompanied by large detunings and the result is a small, but  nonzero dispersive shift on the cavity.  This enables readout when $g_{10-00} = 0$.

In multiqubit gate operations, the finite couplings to these levels will contribute to each qubit’s Stark shift and result in potentially unwanted phase shifts and residual ZZ coupling~\cite{DiCarlo2009}. This accumulated phase can be corrected for by using a qubit refocusing pulse, as in NMR experiments~\cite{Jones1999}. It should also be possible to bias the qubit in a regime where the phase shift is reduced at the expense of the isolation to off-resonant Rabi driving.  Here, we have instead biased for improved single qubit gate operations, where the ability to tune $g_{01-00}$ would reduce crosstalk errors. Other potential uses include a reduction of the Purcell decay rate, a solution for spectral crowding, and a path towards high fidelity, single shot measurement~\cite{Gambetta2011,Filipp}. The long coherence times and ability to coherently manipulate the TCQ’s coupling and frequency on a fast time scale make it a potentially useful device for experiments in quantum optics and quantum computing.

\begin{acknowledgments}
	We thank Blake Johnson, Jerry Chow, and David Schuster for helpful conversations.  This work was supported in part by IARPA under contract W911NF-10-1-0324 and ARO under contract W911NF-11-1-0086.
\end{acknowledgments}

\bibliography{Dispersive2}

\end{document}